\documentclass[lnbip,natbib,sort&compress,numbers,square]{svmultln}
\usepackage{paralist}
\usepackage{makeidx}  % allows for indexgeneration
\usepackage{listings}
\usepackage{graphicx}
\usepackage{calc}
\usepackage{caption}
\usepackage{color}
\usepackage{booktabs}
\usepackage{tabularx}
\usepackage{multirow}
\usepackage{amsmath}
\usepackage{subcaption}
\usepackage{pifont}
\usepackage{amssymb}
\newcommand{\xmark}{\ding{53}}%
\renewcommand{\subsubsection}[1]{\smallskip\noindent{\textbf{#1}}}
\renewcommand{\paragraph}[1]{\smallskip\noindent{\textit{#1}}}
\newcommand{\vspaceSecBefore}[0]{\vspace{-.7em}}
\newcommand{\vspaceSecAfter}[0]{\vspace{-.4em}}
\newcommand{\vspaceSSecBefore}[0]{\vspace{-.6em}}
\newcommand{\vspaceSSecAfter}[0]{\vspace{-.1em}}

\PassOptionsToPackage{hyphens}{url}\usepackage[pdftex, colorlinks=true, hyperfootnotes=false, hyperindex=true, plainpages=false, pagebackref=false, pdfpagelabels=true, pdfstartview=FitH, linkcolor=blue, citecolor=blue, urlcolor=blue, bookmarks=false]{hyperref}
\usepackage[capitalise,nameinlink]{cleveref}

\begin{document}
\mainmatter             
% start of the contribution
%
% \title{Process Channels: Blockchain-based State Channels as a New Layer for Process Enactment}
\title{Process Channels: A New Layer for Process Enactment Based on Blockchain State Channels}
% \title{Process Channels: Enacting Processes on Scalable Blockchain-Based State Channels}
%
\titlerunning{Process Channels: A New Layer for Process Enactment}  % abbreviated title (for running head)
%                                     also used for the TOC unless
%                                     \toctitle is used
%
\author{Fabian Stiehle\inst{1} \and Ingo Weber\inst{1,2}
% Ingo Weber\inst{1 ,2} \orcidID{0000-0002-4833-5921}
}
%
%\authorrunning{Ivar Ekeland et al.}   % abbreviated author list (for running head)
%
%%%% list of authors for the TOC (use if author list has to be modified)
%\tocauthor{Ivar Ekeland, Roger Temam, Jeffrey Dean, David Grove,
%Craig Chambers, Kim B. Bruce, Elisa Bertino}
%
\institute{
% IT Service Management, Development and Operations, Technical University of Munich, Germany\\
% \email{fabian.stiehle@tum.de, ingo.weber@tum.de}
Technical University of Munich, School of CIT, Germany, \email{first.last@tum.de} \and
    Fraunhofer Gesellschaft, Munich, Germany
}

\maketitle              % typeset the title of the contribution
% \index{Ekeland, Ivar} % entries for the author index
% \index{Temam, Roger}  % of the whole volume
% \index{Dean, Jeffrey}
\begin{abstract}
For the enactment of inter-organizational business processes, blockchain can guarantee the enforcement of process models and the integrity of execution traces. However, existing solutions come with downsides regarding throughput scalability, latency, and suboptimal tradeoffs between confidentiality and transparency.
To address these issues, we propose to change the foundation of blockchain-based process enactment: from on-chain smart contracts to \textit{state channels}, an overlay network on top of a blockchain. 
State channels allow conducting most transactions off-chain while mostly retaining the core security properties offered by blockchain. 
%Our approach, process channels, is centered around a protocol for enacting processes on state channels and aims to retain the desired blockchain properties while reducing the on-chain footprint as much as possible. 
Our proposal, \textit{process channels}, is a model-driven approach to enacting processes on state channels, with the aim to retain the desired blockchain properties while reducing the on-chain footprint as much as possible. 
We here focus on the \textit{principled approach of state channels as a platform}, to enable manifold future optimizations in various directions, like latency and confidentiality.
We implement our approach prototypical and evaluate it both qualitatively (w.r.t.\ assumptions and guarantees) and quantitatively (w.r.t.\ correctness and gas cost).
In short, while the initial deployment effort is higher with state channels, it typically pays off after a few process instances---considerably reducing cost.
And as long as the new assumptions hold, so do the guarantees.
\keywords{Blockchain, Business Process Execution, Choreographies, Interorganisational processes, State Channels}
% \keywords{Blockchain \and Choreographies \and Business Process Enactment \and Business Process Execution \and State Channels}
\end{abstract}
\vspaceSecBefore
\section{Introduction}
\label{sec:intro}
\vspaceSecAfter
For the enactment of inter-organizational processes, blockchain can guarantee the enforcement of rules and the visibility and integrity of execution traces---without introducing a centralised trusted party. 
The current state of the art focuses on on-chain enactment, where a process model is transformed into a smart contract and executed on the blockchain~\cite{stiehle2022blockchain}. 
However, blockchain execution comes with downsides and suboptimal tradeoffs regarding scalability and confidentiality.~\cite[Chapter~3]{xuArchitectureBlockchainApplications2019}.
On-chain process enactment inherits these problems; to address this, related work has focused on improving the cost of the on-chain components (e.g.,~\cite{garcia-banuelosOptimizedExecutionBusiness2017a, lopez-pintadointerpretedExecutionBusiness2019,  loukilDecentralizedCollaborativeBusiness2021a}). 
In contrast, we propose a more fundamental change: to move from full on-chain enactment to layer two state channels. 
Layer-two technologies have emerged as a promising direction to address fundamental challenges of blockchain technology~\cite{gudgeonSokLayertwoBlockchain2020}. One of these technologies are state channels. State channels move the bulk of transactions into off-chain channels. In these channels, participants transact directly without the involvement of the blockchain. The blockchain is only used for channel creation and as a dispute resolution and settlement layer. This can greatly reduce the on-chain footprint, enabling new levels of scalability and improved confidentiality, while mostly retaining the core security properties offered by blockchain. 
%New levels of scalability, in turn, can enable manifold future opportunities, such as the use of privacy-preserving technologies, so far deemed too costly~\cite{zhang2019zchannel}.
 
In this paper, we focus on conceptualising the principled approach of enacting processes in state channels. We focus on the fundamental aspects of this new approach: conceiving how to achieve the main functionality and quality attributes, studying where advantages and disadvantages lie, and creating a basis for a new line of research.
To this end, we propose \textit{process channels}: a model-driven approach to transform process models into state channel constructions. 
%As part of it, we present a protocol for enacting choreographies in channels. 

To evaluate our approach, we develop a prototype: \textit{Leafhopper}. As we propose a new platform for blockchain-based enactment, we provide a qualitative assessment and investigate the guarantees and assumptions compared to on-chain enactment. We also provide a quantitative evaluation and benchmark our prototype in terms of correctness and gas cost. While gas cost is foremost known as a measure for transaction cost on Ethereum, it is directly linked to the amount, size, computational complexity, and storage requirement of transactions~\cite{buterinEthereum}---serving well as a measure to assess the on-chain footprint.
We find that, while the initial deployment cost of Leafhopper is higher, it can considerably reduce gas cost. 
% For our investigated cases, the higher deployment cost pays off after the execution of 3 process instances. Furthermore, while the cost of on-chain enactment corresponds to the complexity of the process, Leafhopper exhibits different cost factors. The most impactful is the likelihood of an on-chain dispute. Taken typical contract dispute rates from industry, Leafhopper vastly reduced gas cost compared to on-chain enactment.
Leafhopper does so without weakening the main security guarantees of the blockchain, given that some additional assumptions hold. 
% Beyond our proposal, we picture future developments made possible by process channels.

Following open science principles, we make the entire code and data of our prototype and evaluation publicly available -- see Footnote~\ref{fn:repos}. Beyond Leafhopper, we also publish \textit{Chorpiler}, the first open-source compiler capable of generating optimised smart contracts from choreographies.
In the remainder of the paper, we first discuss the background (\cref{sec:background}) and related work (\cref{sec:relw}), before presenting the process channel approach (\cref{sec:approach}). Implementation and evaluation are covered in \cref{sec:eval}, before \cref{sec:concl} concludes.
\vspaceSecBefore
\section{Blockchain and Layer Two Channels}
%\section{Background}
\label{sec:background}
\vspaceSecAfter
% TODO: kuerzen auf key points: guarantees of blockchain, types of blockchain, problems of types of blockchain
%\subsection{Blockchain}
A blockchain is an append-only store of transactions distributed across a network of nodes~\cite[Chapter~1]{xuArchitectureBlockchainApplications2019}. This data store is called ledger; the ledger and the network of nodes form the defining parts of a particular blockchain system.
Participants in the network are identified through their blockchain address, which is derived from a public key. Each transaction is cryptographically signed with the sender's corresponding private key, and is validated according to the blockchain system’s protocol. Smart contracts allow to execute user-defined programs on the blockchain. In practice, a blockchain system can provide integrity, immutability, non-repudiation, equal-rights, and full transparency~\cite[Chapter~1.4]{xuArchitectureBlockchainApplications2019}
%We can distinguish between public, private, and permissioned blockchains. A public blockchain comprises an open network in which each node can propose and verify new blocks. In contrast, a private blockchain is not openly accessible. A permissioned blockchain includes a layer that defines membership rules, e.g., who can propose new transactions. 
%Usually, a private or permissioned blockchain can operate more efficiently, as they do not have to protect against sybil attacks; however, this comes at the cost of reduced decentralisation~\cite[Chapter~3.2]{xuArchitectureBlockchainApplications2019}. 
%
%\subsection{Layer Two Channels}
%

Layer two channels attempt to scale the underlying blockchain layer by offloading transactions~\cite{gudgeonSokLayertwoBlockchain2020}. The blockchain is no longer involved in every minute transaction---these are moved into channels. The idea first emerged in the concept of payment channels (e.g.,~\cite{poonBitcoinLightningNetwork2016}). 
Say, you want to pay an online news site \$0.10 per article that you read. You create a channel with the news site, where you lock \$5.00 as initial funds (or \textit{collateral}). Every time you read an article, you exchange an off-chain transaction with the news site, assigning an additional \$0.10 to their account. After 32 articles, you decide to close the channel, with the accrued \$3.20 assigned to the news site and the remaining \$1.80 refunded to your account.
This concept has since been generalised to state channels~\cite{millerSpritesStateChannels2019}.
Participants wishing to transact first agree on a contract governing the rules of the channel and encode these on the blockchain (e.g., in a smart contract). They then conduct off-chain transactions directly across the channel. For each transaction, they agree on the outcome and cryptographically commit to their agreement.
Finally, when they have concluded their contract, they submit the final state to the chain.
If at any point a participant (supposedly) violates the rules, e.g., attempts to falsify the outcome of a transaction, or become unavailable, the last unanimously agreed transaction is posted to the blockchain. From this state, participants can then safely resume their interaction on the chain, where the blockchain protocol enforces the  honest execution of the contract. 
%In the best case, a contract can be completed successfully off-chain. At any time participants can be convinced, that the blockchain will accept a previously agreed state change. 

More specifically, a channel constructs off-chain peer-to-peer connections between all channel participants. The off-chain channel replicates an on-chain state machine, e.g., a smart contract, called the \textit{channel contract}. Both can be modelled with state $i$ and state transition function $step$. The transition function takes a set of commands $cmd_{i+1}$ and transitions the state from $i$ to $i+1$. 
%it computes: $state_{i+1}\leftarrow step(state_i, cmd_{i+1})$.  
Channels typically transition through the following lifecycle phases~\cite{gudgeonSokLayertwoBlockchain2020}:
\begin{enumerate}
\item\textbf{Establishment}: All participants agree on a channel contract, which encodes the rules of the channel and the initial state. Usually, this phase involves on-chain activity, such as locking collateral or deploying the contract.

\item \textbf{Transition(s)}: A participant proposes a new state transition $step_{i+1}$ through a cryptographically signed message. Every other participant verifies that $step_{i+1}$ is in conformance with their local version of the contract. If so, they confirm by signing it and sending their signature to every other participant. Once a participant receives all other signatures on transition $i+1$, they can consider the result of $step_{i+1}$ the new valid state.

\item \textbf{Dispute}: If a participant did not receive all signatures on a state transition in time, or they receive conflicting transitions, they must assume that some fault occurred. They must now start a dispute process on the blockchain. To do so, the participant submits the last unanimously signed state transition to the channel contract, which will install the result as new current state. 
This closes the off-chain channel and starts the dispute window. Should a participant submit a state transition that has already been superseded by a more recent one, so called \textit{stale state}, another participant must notice and submit the most recent state. 

\item \textbf{Closure}: After a channel closes and the dispute window elapses, future transitions can be sent directly to the channel contract, which ensures the continued validity of transitions.  A channel can also be closed by unanimous vote, installing the last agreed state transition as final state of the contract.
\end{enumerate}
Channels introduce new assumptions that must hold~\cite{gudgeonSokLayertwoBlockchain2020}:
\begin{enumerate}
\item \textbf{Blockchain Reliability}: The security properties of the blockchain hold and valid transactions submitted to the blockchain are eventually accepted. 

\item \textbf{At Least One Honest Participant} must be present in the channel to contest faulty state transitions.

\item \textbf{Always Online}: State channels require participants to remain online during the entire lifecycle of the channel to prevent execution forks~\cite{mccorryPisaArbitrationOutsourcing2019}, in which a malicious actor starts the dispute phase and submits stale state to the blockchain, e.g., $state_{i-1}$. Honest participants must notice such an attempt and submit $state_i$.\footnote{We generally assume a party ``looks after themselves", and follows a strategy with the highest payoff.}
\end{enumerate}
Given these assumptions, channels can achieve safety: the integrity of the contract's state is ensured, even when all parties but one are malicious; and liveness: an honest participant can always advance the contract given a valid transition function, even when all other parties try to stall the process~\cite{millerSpritesStateChannels2019}.

\vspaceSecBefore
\section{Related Work}
\label{sec:relw}
\vspaceSecAfter
The current state of the art in blockchain-based business process enactment focuses on on-chain enactment~\cite{stiehle2022blockchain}. %Scalability, in terms of cost, throughput, and latency remain an issue. %Following the seminal work of Weber et al.~\cite{weberUntrustedBusinessProcess2016a}, the research goal has been to improve the capabilities (eg.,~\cite{lopez-pintadoCaterpillarBusinessProcess2019a, luIntegratedModelDriven2021a, lopez-pintadoControlledFlexibilityBlockchainbased2022a, abidModellingExecutingTimeAware2020}) support other modelling methodologies (eg.,~\cite{meroniTrustedArtifactDrivenProcess2019, madsenCollaborationAdversariesDistributed2018a}), or demonstrate improved cost of the on-chain components (eg.,~\cite{lopez-pintadointerpretedExecutionBusiness2019, sturmLeanArchitectureBlockchain2018, loukilDecentralizedCollaborativeBusiness2021a}).
%In a recent survey, we reported that scalability in particular remains an open challenge~\cite{stiehle2022blockchain}.
To reduce the on-chain footprint, related work has focused on improving the cost of the on-chain components. Garc{\'i}a-Ba{\~n}uelos et al.~\cite{garcia-banuelosOptimizedExecutionBusiness2017a} introduced an optimised generation of smart contracts through petri net reduction. We extend this approach for BPMN choreographies. 
% Sturm et al.~\cite{sturmLeanArchitectureBlockchain2018}, 
L{\'o}pez-Pintado et al.~\cite{lopez-pintadointerpretedExecutionBusiness2019}, and Loukil et al.~\cite{loukilDecentralizedCollaborativeBusiness2021a} propose an interpreted approach, where a process model is not compiled but executed by an interpreter component on the blockchain. While the deployment becomes more costly, it leads to cost savings over multiple instance runs. 

In a recent survey, we pointed to layer two technologies as a promising research direction~\cite{stiehle2022blockchain}.
To the best of our knowledge, this is the first work to investigate layer two state channels for enacting business processes.
The state of the art in state channel\footnote{Hyperledger Fabric uses the terminology of channels for their subnet functionality~\cite{androulaki2018hyperledger}. The similarity to state channels is weak; like subnets, fabric channels partition the on-chain ledger. State channels construct off-chain channels and use the security guarantees of the on-chain ledger as settlement and dispute resolution layer.} constructions focuses on the formalisation and security of protocols for general applications (e.g.,~\cite{millerSpritesStateChannels2019,dziembowskiGeneralStateChannel2018, dziembowskiMultipartyVirtualState2019}). More in line with our work, McCorry et al.~\cite{mccorryYouSankMy2019} present a case study investigating the feasibility and applicability of their state channel construction. They present a template for migrating an existing smart contract to a state channel construction. While they present a state channel architecture for \textit{n} parties, they chose a two-party game as their case. In two-party games, participants can take turns. However, when more participants are involved, a schedule becomes necessary~\cite{dziembowskiMultipartyVirtualState2019} (see also \cref{sec:schedule})
We present process channels:
%\footnote{In the early days of choreography languages, languages like WS-CDL (CITE) were based on the formal $\pi$-calculus (CITE), which uses a concept of channels for structuring communication. The relation between our work and these early approaches is weak. Notable differences include (i) the possibility of more than two participants per channel, and (ii) the level of abstraction: our models being executable, whereas WS-CDL and $\pi$-calculus are much more abstract, the latter even abstracting from branching conditions.}
a model-driven approach, where a process model is used to automatically generate the entire channel setup. The process control-flow naturally enforces a schedule upon the participants. We discuss the particularities of enforcing choreographies in channels and evaluate our approach quantitatively and qualitatively in comparison to a full on-chain baseline. For our evaluation, we chose two commonly used business processes, which allows us to compare our approach to existing related work.

\vspaceSecBefore
\section{Process Channels}
\label{sec:approach}
\vspaceSecAfter
In this section, we first give an overview of our approach. We, then, address a particular challenge when designing \textit{n}-party state channels: the scheduling problem. After, we can describe our model transformation approach, and, lastly, outline the channel's protocol in detail. 
\vspaceSSecBefore
\subsection{Overview}
\vspaceSSecAfter
\begin{figure}[tb]
    \centering
    %\vspace{-20pt}
    \includegraphics[width=.85\linewidth]{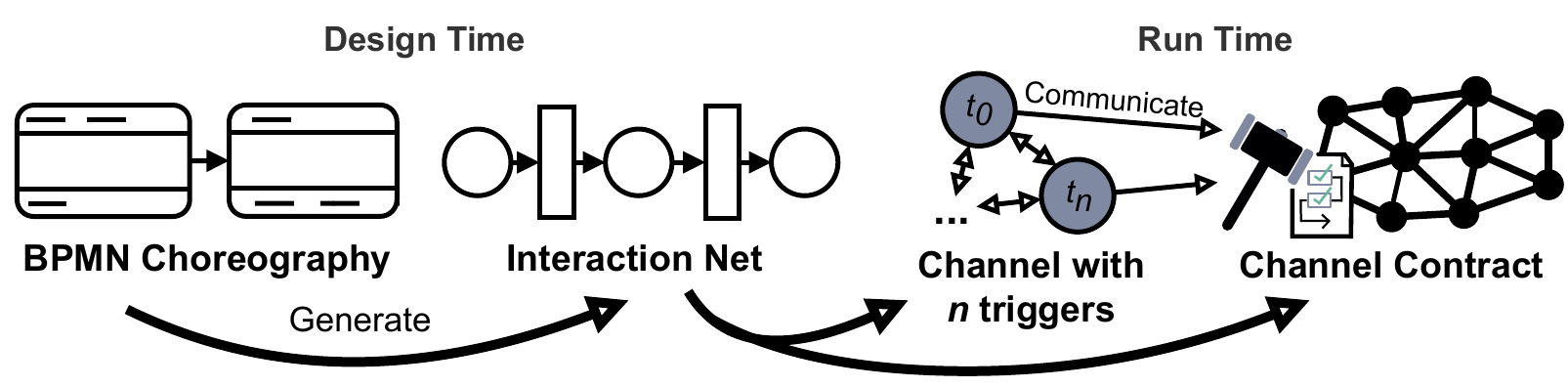}
    \caption{Overview: From BPMN choreographies to process channels.}
    \label{fig:my_label}
    %\vspace{-20pt}
\end{figure}
In Figure 1, we show an overview of our approach. At its core is the model-driven engineering paradigm. This paradigm has been found to address many challenges of developing blockchain-based applications, specifically, reduce blockchain specific complexity~\cite{diciccioBlockchainSupportCollaborative2019}. Channel constructions exacerbate this complexity issue further. More functionality is required, both on and off the chain, thus, also introducing a new trust concern: the security of the channel protocol. To alleviate this issue, we propose to generate channel components from a process model. 
We base our approach on
BPMN choreography diagrams. %, which are an abstraction of choice for modelling process choreographies~\cite{OMG_BusinessProcessModel_2011}. 
These fit well the model of state channels, where the focus is on a set of interconnected autonomous participants, initiating messages according to some schedule. 
From a BPMN choreography model, a compiler component generates an interaction Petri net. This serves as a middle layer presentation, and allows us to apply optimisation techniques before compilation. From the optimised interaction net, we generate the \textit{channel trigger} components.
Furthermore, the process channel contract is generated, which is deployed on the blockchain. 
Each participant deploys a trigger and exposes it to the channel network. 
Each trigger is interconnected. Additionally, each trigger must be connected to the blockchain, providing access to the channel contract. We assume that each organisation runs a process-aware information system (PAIS) in their (private) organisational network. The PAIS communicates with the trigger to interact with the network. In particular, the responsibilities of each generated component are as follows.
\begin{itemize}
    \item The \textbf{channel contract} handles channel establishment, disputes, and closure, and is deployed on the blockchain. Upon deployment, the blockchain addresses of the participants are bound to their corresponding role. Should a dispute be triggered, the contract validates the submitted state transition by verifying that it has been signed by all participants. The channel also contains process enactment capabilities: it enforces the honest continuation of the contract, should a dispute have occured. 
    \item The \textbf{channel trigger} communicates with the channel network to enact the process model. Each trigger must be configured with: the identify and secret key of its participant, the  blockchain addresses of the other participants, the host information of the other triggers, and the address of the channel contract.
    Once a trigger receives a request from its PAIS, the trigger performs a conformance check to verify the request locally; when successful, it proposes a new state transition to the network. The trigger monitors continuously whether the channel contract has transitioned into the dispute phase. Should a trigger not be able to advance the process, or receive a non-conforming transition request, it starts a dispute phase invoking the channel contract.
\end{itemize}
\vspaceSSecBefore
\subsection{The Scheduling Problem}
\label{sec:schedule}
\vspaceSSecAfter
In an \textit{n}-party state channel construction, multiple concurrent proposals can deadlock the protocol, where a subset of participants is waiting for the consent to a proposed state transition, while other participants, in turn, are waiting for the consent of a concurrently proposed transition~\cite{dziembowskiMultipartyVirtualState2019}. In 2-party state channels, participants can simply take turns (e.g., in \cite{dziembowskiGeneralStateChannel2018}). However, for \textit{n}-parties, this problem constitutes a leader election problem. A probable solution is the utilisation of a leader election algorithm; however, this would introduce communication overhead and is not done in practice. Instead, this problem is either not addressed in literature, or a specific schedule of turns is enforced over all participants (see \cite{negka2021blockchain} for a survey). However, the scheduling of a process is a well understood problem in the world of business process management. A process model inherently contains rules considering the order of events, while a process choreography contains rules regarding the roles and interactions of participants~\cite[Chapter~6]{weskeBusinessProcessManagement2019}. 
%furthermore, they can be investigated regarding their compatibility~\cite[Chapter~6]{weskeBusinessProcessManagement2019}. 
We can make direct use of this control-flow information to derive a valid schedule to be enforced within the state channel.
\vspaceSSecBefore
\subsection{Optimised Generation}
\label{sec:generation}
\vspaceSSecAfter
As outlined, channel contract and trigger components require process enactment capabilities. We generate these from a BPMN choreography model.
Our approach, hereby, is based on the optimised translation technique presented in {Garc{\'i}a-Ba{\~n}uelos} et al.~\cite{garcia-banuelosOptimizedExecutionBusiness2017a}: a process model is converted into a Petri net, this net is then reduced according to well-established equivalence rules. From the optimised net, code is generated. In the code, the process state is then encoded as efficient bit array.
While~\cite{garcia-banuelosOptimizedExecutionBusiness2017a} is based on BPMN process models, we use BPMN choreography models. Thus, our approach is based on interaction Petri nets, which are a special kind of labelled Petri nets. Interaction Petri nets have been proposed as the formal basis for BPMN choreographies~\cite{decker2007local}. As labels, they store the initiator and respondent information, which are essential for the channel transitions. After conversion, we apply the same reduction rules as in~\cite{garcia-banuelosOptimizedExecutionBusiness2017a}. For this contribution, we limit the scope to choreography tasks, start and end events, and parallel and exclusive gateways. As in~\cite{garcia-banuelosOptimizedExecutionBusiness2017a}, this also supports looping behaviour. In contrast to~\cite{garcia-banuelosOptimizedExecutionBusiness2017a}, we must restrict enforcement to certain roles: only initiators are allowed to enforce tasks.\footnote{A choreography task can be one-way or two-way: i.e., it optionally includes a response. We assume that a choreography task is one-way; two-way tasks can be regarded as syntactic sugar and adding support for those is no conceptual challenge.} Here, we can differentiate between \textit{manual} and \textit{autonomous transitions}. Manual transitions correspond to tasks that are initiated by a participant; these must be explicitly executed. Autonomous transitions are the remaining silent transitions.
Converting a process model into a Petri net creates silent transitions, and while most of them can be deleted through reduction, some cannot be removed without creating infinite-loops~\cite{garcia-banuelosOptimizedExecutionBusiness2017a}. These transitions must then be performed by the blockchain autonomously, given that the correct conditions are met. Consequently, these transitions are not bound to a role. The differentiation allows a more efficient execution: if the conditions for a manual task are met, it is fired and terminated; further autonomous transitions may be fired, without requiring further manual transitions.
\vspaceSSecBefore
\subsection{Channel Protocol}
\label{sec:protocol}
\vspaceSSecAfter
Once the components are generated, they execute the channel protocol. 
In the following, we outline the protocol of our channel construction, based on the channel lifecycle model introduced in \cref{sec:background}.

\subsubsection{Establishment.}
For the following, we assume all channel triggers are deployed and have established, secure connections. From here, blockchain addresses are exchanged between participants. Any participant can now deploy the channel contract. The deployment initialises the contract, binding all addresses to their role and setting the initial state. The address of the deployed contract is then distributed to all participants; these verify the contract to ensure it was initialised with the correct addresses and the correct initial process state.\footnote{This procedure can be made easier by forcing deployment from an agreed upon channel factory contract~\cite[Chapter~7.4.4]{xuArchitectureBlockchainApplications2019}.} If the contract passes verification, the contract is accepted as channel contract.
\begin{figure}[tb]
    \centering
    \includegraphics[width=.8\linewidth]{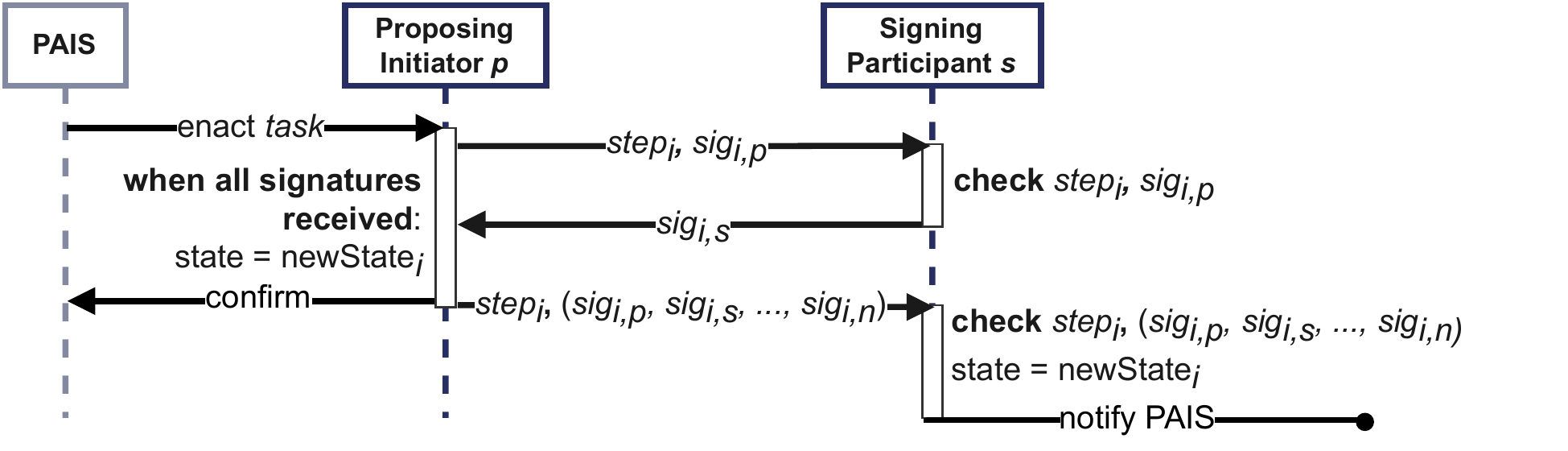}
\caption{Sequence diagram for a channel transition (happy path).}
\label{fig:transition}
\vspace{-5pt}
\end{figure}

\subsubsection{Transition.}
We depict the protocol for a state transition for a sequence flow from state $state_{i-1}$ to $state_{i}$ in \cref{fig:transition}. The PAIS sends an enactment request for a task to its corresponding channel trigger. The task must encode all required information to compute the new process state and can  additionally include arbitrary data.
The trigger verifies that the enactment of the task is in conformance with its local process state. It then becomes the \textit{proposing initiator} $p$ for this task, and prepares the state transition proposal $step_i$. Each transition proposal is assigned a sequence number $i$, which is incremented after each successful transition. $step_i$ includes the proposed task and resulting state $newState_i$.\footnote{To prevent the replay of transitions across cases, instances, or blockchains, unique identifiers must also be included, e.g., case ID, instance ID, and chain ID.} The trigger cryptographically signs the proposed transition and sends $step_i$, and its signature $sig_{i,p}$ to all other participants, called the \textit{signing participants}. All signing participants verify that $step_i$ was proposed by the correct initiator by verifying the signature, and that $step_i$ leads to the next conforming state. If all checks pass, each signing participant $s$ signs the new step and sends their signature $sig_{i,s}$ back to the initiator. Once the initiator has collected all signatures $(sig_{i,s}, ..., sig_{i,n})$, it can accept $newState_i$ as new state of the process. It now confirms the transition proposal by sending the set of signatures to all signing participants. These also verify the signatures and, when all checks pass, can also accept $newState_i$. All participants must store the received signatures and corresponding transition proposals, as they are required should a dispute occur.\footnote{To reduce the amount of messages, confirmations can be prepended to a transition proposal. That is, once an initiator has collected all signatures for $step_{i}$., it only sends the confirmation to the next initiator. The next initiator prepends the confirmations to the next transition proposal $step_{i+1}$.}

As we have discussed in \cref{sec:schedule}, a problem of \textit{n}-party state channels are multiple concurrent state transition proposals. Imposing an order of transitions is, therefore, paramount. Using the control-flow information of the model, it is often trivial to enforce such an order. Consider, a simple sequence flow from task $t$ to task $t'$. The ordering $t < t'$ naturally follows. This is less so when control-flow branches after gateways. For exclusive gateway branches, it suffices to collapse the possible branches into one during run time. That is, the conditions present on the outgoing sequence flows of the gateway must be based on internal channel state and be part of the transition proposal, so it can be made available to the blockchain in case of a dispute; it can not be based on external state. This requirement can be lifted when all immediately following tasks belong to the same initiator, as it then becomes a (private) choice of a single participant.

Parallel gateways, on the other hand, permit concurrent behaviour and a unique order is not enforceable. Real concurrency is a non-trivial problem in state channel constructions~\cite{negka2021blockchain}.
As each state transition must encode a sequence number and the new global process state, a concurrent execution can deadlock the off-chain protocol. We, thus, require that tasks on parallel branches are serialised at run time into any order chosen by the participants. 
Should a deadlock still occur, the blockchain contract can always enforce an order during a dispute phase, as the blockchain ledger enforces a unique order of transactions.
%To illustrate this, let us for a moment consider that concurrent execution would be allowed for parallel gateways. Consider, the process excerpt in Figure 3b. Consider the last confirmed transition in the channel to be $step_A$, terminating task $t_A$. Now consider a participant proposing $step_B$, leading to a new process state, with the termination of task $t_B$, and task $t_C$ still enabled. Consider a concurrent proposal $step_C$, which will lead to a different process state with task $t_C$ terminated, and task $t_B$ still enabled. As both transitions are considered valid, both will be accepted and signed by the participants. However, these two transitions cancel each other out, as one leaves task $t_B$ enabled and the other leaves $t_C$ enabled. In a dispute phase, the blockchain would accept both transitions, given the correct signatures, and given that they can be submitted in any order. This would lead to a situation where both participants can repeatedly submit $step_B$ and $step_C$, cancelling each other out. Thus, for branching behaviour, the contract also has to allow branching state.
%We leave this to future work. For this contribution, participants have to agree on any sequential order of parallel tasks. If they can not agree on an order, the blockchain can resolve the dispute by relying on the totally ordered ledger of transactions. That is, the contract will accept the order in which the transitions are appended to the ledger, i.e., on a first come basis.

\subsubsection{Dispute.}
At any point, the channel contract is in a certain state $i$. A participant can trigger a dispute by providing all participants’ signatures for state transition $step_j$, where $i < j$. Once the contract is in dispute mode, additional state can be submitted until the dispute window elapses. Then, the contract continues on-chain. We describe possible dispute scenarios. 
\begin{itemize}
    \item \textit{Non-Conforming Transition}: Consider a non-conforming transition is proposed in the channel. Assuming there is at least one honest participant, the transition would not acquire unanimous consent. A participant can trigger a dispute submitting the last agreed transition to the blockchain; thus, forcing the continuation of the contract on the blockchain. Faulty participants can, however, also stop to take part in the protocol, which we discuss next.
    \item \textit{Unavailability:} Consider a transition is proposed to the channel. After a local timeout, the initiator does not receive signatures from all participants. To ensure liveness, a dispute must be triggered to force the transition on-chain. Consider the reverse: after a local timeout, a signing participant does not receive the expected transition proposal or confirmation. The participant can trigger a dispute. Now, the initiator has to perform the transition on-chain or be identified as participant stalling the process, which could be penalised.
\end{itemize}

\subsubsection{Closure.}
Once participants reach the end event in unanimous consent, they submit the final state to the channel contract. Otherwise, the end event is reached on-chain. In both scenarios, the process terminates and a new case can be instantiated. The contract assigns a new case ID and resets the process state.

\vspaceSecBefore
\section{Implementation and Evaluation}
\label{sec:eval}
\vspaceSecAfter
To enable the evaluation of our approach, we developed two prototypes, \textit{Chorpiler} and \textit{Leafhopper}. We perform a quantitative and qualitative evaluation and compare our approach to an on-chain enactment baseline. 
For the quantitative evaluation, we use process models from well known cases from literature.
%, enabling comparison with approaches from the related work.
We verify the correctness of our implementation by replaying process traces and perform benchmarks to assess cost. For the qualitative evaluation, we discuss the required assumptions and provided guarantees of process channel enactment in comparison to full on-chain enactment.
\vspaceSSecBefore
\subsection{Implementation and Setup}
\vspaceSSecAfter
\subsubsection{Chorpiler and Leafhopper}.
We have developed two tools, Chorpiler and Leafhopper.
Chorpiler implements the optimised generation of enactment components, as outlined in \cref{sec:approach}. It is capable of generating process channel contracts and on-chain enactment contracts in Solidity, as well as enactment functionality in TypeScript, which is used in the channel triggers.
The static trigger component capabilities, e.g., routing, signature verification etc., are implemented in Leafhopper and run on \textit{Node.js}\footnote{See \textit{Node.js}, \url{https://nodejs.org/en}, accessed 2023-03-17.}. Leafhopper uses Chorpiler to generate the process channel contract and the enactment capabilities of the channel trigger. For each participant in the choreography, a trigger is deployed. 
For ease of deployment, each trigger is run in a \textit{Docker} container and the trigger network can be deployed using \textit{Docker Compose}.\footnote{%See Docker, \url{https://www.docker.com} and Overview of 
See \textit{Docker Compose}, 
\url{https://docs.docker.com/compose}, %both 
accessed 2023-03-17.} 

\subsubsection{Benchmark Setup}. We benchmark the supply chain~\cite{weberUntrustedBusinessProcess2016a} (adapted from~\cite{fdhila2015changeandcompliance}) and incident management~\cite{OMG2010BPMNbyExample} case, which are well known from related work.
To help assess our approach, we compare it to a baseline. The baseline provides the same model support as Leafhopper, but enacts the process completely on-chain, as in related work. For each case, we generate the baseline, channel triggers, and channel contract. The triggers are run in a local network. The smart contracts are deployed to an Ethereum environment; we use the Ethereum simulation tool \textit{Ganache}.\footnote{See \textit{Ganache}, \url{https://trufflesuite.com/ganache}, accessed 2023-03-17.}
Following open science principles, and to enable replicability, we made both our prototypes,  evaluation scripts, and data publicly available.\footnote{\label{fn:repos}Leafhopper is available at \url{https://github.com/fstiehle/leafhopper}. The repository includes instructions and scripts to automate the replication of our evaluation. Chorpiler is available at \url{https://github.com/fstiehle/chorpiler}.}
\vspaceSSecBefore
\subsection{Quantitative Evaluation}
\label{sec:quantitative}
\vspaceSSecAfter
\subsubsection{Correctness.}
We verify that the network only accepts conforming traces and always remains in a stable state, i.e., all triggers report the same state after some finite time. To do so, we follow the methodology outlined in~\cite{weberUntrustedBusinessProcess2016a}. For each case, we replayed all conforming traces (two for supply chain and four for the incident management case). After, we generated 2000 non-conforming traces; to do so, a conforming trace was randomly manipulated by one of the following operations: add an event, remove an event, and swap the position of two events.\footnote{We removed any coincidentally created conforming traces. In total we replayed $1812$ non-conforming traces to the incident mgmt.\ and $1933$ to the supply chain case.} We replayed these traces from a local script which, for each event, connects to the corresponding initiator to propose the task.\footnote{Normally, the local trigger would also verify the request and only forward valid requests. We disabled this functionality to allow us to simulate a faulty component.} All events were classified correctly w.r.t.\ conformance, and after each trace replay the channel was in a stable state. 

\subsubsection{Cost Analysis.}
We compare the cost of our baseline to Leafhopper. 
For the baseline, we replayed, for each case, all conforming process variants (two for supply chain and four for the incident management case) and recorded the gas cost of all interactions with the blockchain. As gas costs are deterministic, multiple runs of the same variant are not required.

\noindent Cost in Leafhopper is more difficult to assess and is driven by the cost for the channel establishment (deployment of the contract) and successful closure (unanimous submission of the final state) or dispute.
To study the cost behaviour, %For the channel implementation, 
we performed, analogous to our baseline, for each conforming process variant the following benchmarks:
\begin{enumerate}
    \item A \textbf{best case} run with no disputes, where the channel is unanimously closed.
    \item A \textbf{bad case} run, where a dispute is triggered after half of the process. As a result, the other half must be completed on the blockchain.
    \item A \textbf{worst case} run, where a dispute with stale state is made immediately after the start event. An honest participant then submits the last agreed-upon state, and the entire remaining process must be completed on-chain.
\end{enumerate}
\begin{table}[t]
\scriptsize
\centering
\begin{tabular}{|l|lll|lllll|}
\hline
\multirow{3}{*}{\textbf{Case}} &
  \multicolumn{3}{|c|}{\textbf{Baseline}} &
  \multicolumn{5}{|c|}{\textbf{Leafhopper}} \\ \cline{2-9} 
 &
  \multicolumn{1}{|l|}{\multirow{2}{*}{\begin{tabular}[c]{@{}c@{}}Deploy-\\ ment\vspace{-9pt}\end{tabular}}} &
  \multicolumn{2}{c|}{Avg. Exec.} &
  \multicolumn{1}{|l|}{\multirow{2}{*}{\begin{tabular}[c]{@{}c@{}}Deploy-\\ ment\vspace{-9pt}\end{tabular}}} &
  \multicolumn{1}{l|}{\multirow{2}{*}{\begin{tabular}[c]{@{}l@{}}Exec.\\Best\\ Case\end{tabular}}} &
  \multicolumn{3}{c|}{Avg. Exec.} \\ \cline{3-4} \cline{7-9} 
 &
  \multicolumn{1}{|l|}{} &
  \multicolumn{1}{c|}{Case} &
  \multicolumn{1}{c|}{Task} &
  \multicolumn{1}{|l|}{} &
  \multicolumn{1}{l|}{} &
  \multicolumn{1}{c|}{Bad Case} &
  \multicolumn{1}{c|}{Worst Case} &
  \begin{tabular}[c]{@{}c@{}}On-Chain\\ Task\end{tabular} \\ \hline
\begin{tabular}[c]{@{}l@{}}Supply\\Chain\end{tabular} &
  \multicolumn{1}{|r|}{396.732} &
  \multicolumn{1}{r|}{347.076} &
  \multicolumn{1}{r|}{34.708}  &
  \multicolumn{1}{|r|}{772.282} &
  \multicolumn{1}{r|}{88.319} &
  \multicolumn{1}{r|}{310.186} &
  \multicolumn{1}{r|}{495.207} &
  \multicolumn{1}{r|}{38.691}
   \\ \hline
\begin{tabular}[c]{@{}l@{}}Incident\\ Mgmt.\end{tabular} &
  \multicolumn{1}{|r|}{408.954} &
  \multicolumn{1}{r|}{190.509} &
  \multicolumn{1}{r|}{31.752}  &
  \multicolumn{1}{|r|}{784.823} &
  \multicolumn{1}{r|}{88.319} &
  \multicolumn{1}{r|}{199.545} &
  \multicolumn{1}{r|}{328.889} &
  \multicolumn{1}{r|}{34.774} \\ 
  \hline
\end{tabular}
\caption{Gas cost of Leafhopper in relation to the baseline.}
\label{tab:gas_baseline}
\vspace{-5pt}
\end{table}
Table~\ref{tab:gas_baseline} shows the recorded gas costs for our benchmark experiments.\footnote{While our baseline is based on~\cite{garcia-banuelosOptimizedExecutionBusiness2017a}, it incurs increased gas cost (compare with Table~\ref{tab:gas_others}), as it additionally implements role enforcement (c.f. \cref{sec:generation}).} 
Compared to our baseline, Leafhopper incurs around twice the cost for \textit{deployment} due to the implemented channel capabilities. However, the \textit{best case} execution considerably improves upon the avg. execution cost of the baseline. 
Furthermore, we can see that the best case execution cost is fixed. It requires one round of signature verification. The cost, hence, does not depend on the complexity of the process or its tasks, only on the number of participants---five participants for both cases. 

The \textit{bad case} execution cost is still lower for the supply chain case and slightly higher for the incident management case  compared to the avg. execution cost of the baseline. The incident management case is on average shorter (six versus ten tasks of the supply chain case) but has the same number of participants. Thus, the fixed state verification cost has a higher impact on the total cost of the shorter process. 
The \textit{worst case} cost is considerably higher than the baseline's average. This is  expected, as it constitutes two state submissions and the enactment of the entire remaining process on the chain. Should a dispute occur, Leafhopper exhibits slightly (around 10\%) higher cost for enacting a singular task on the blockchain. This is the result of having to determine whether a dispute is currently active.
\begin{figure}[tb]
\centering
  \includegraphics[width=0.8\linewidth]{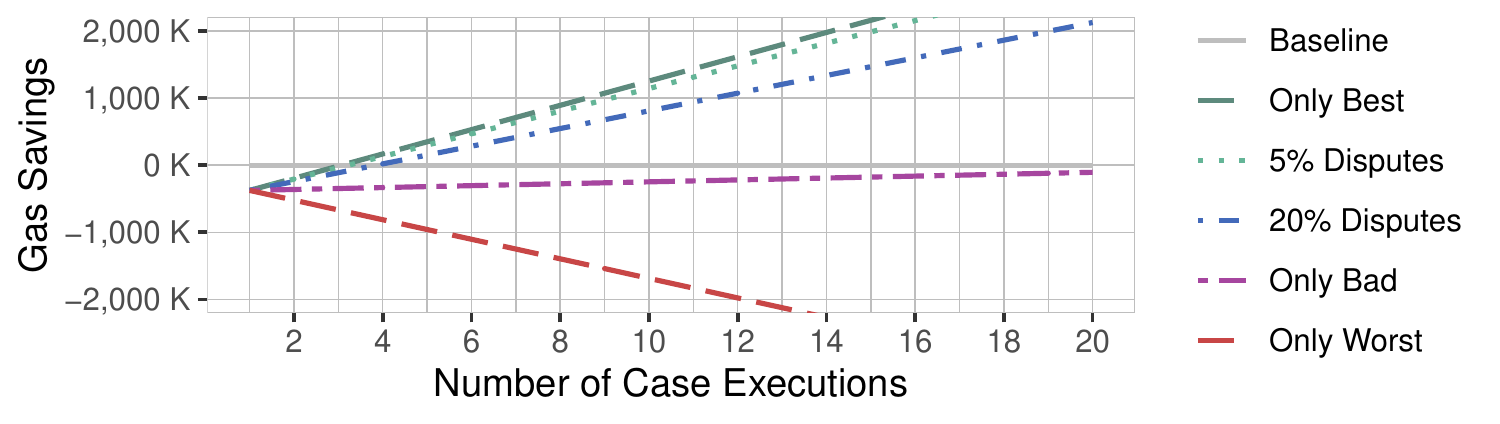}
  \caption{Gas savings compared to baseline, when the initial deployment is re-used.}
  \label{fig:gas_case}
\end{figure}

\paragraph{Cost Under Different Dispute Scenarios.}
To put the cost of Leafhopper into context, we assess the relative overhead of Leafhopper compared to the baseline when considering different dispute rates. That is, the number of expected disputes over multiple case runs. 
We illustrate this in \cref{fig:gas_case}, where we show how the average cost savings for a case execution on Leafhopper develops for different dispute scenarios, when the initial deployment is re-used. 
We depict five scenarios, only best case runs, 5\% disputes, 20\% disputes, only bad, and only worst case runs. For the 5\% and 20\% runs, we assume an equal share of bad and worst cases. 
Both percentages are taken from an industry survey, where 5\% constitutes the best and 20\% the worst measured average contract dispute rate by industry sector.\footnote{IACCM: \textit{Are you in an adversarial industry? Insights for contract negotiators and managers}. 2014. \url{https://wp.me/pa5oX-RH}, accessed 2023-03-28.} 

We can see that, when considering only best cases, the deployment costs of Leafhopper is amortised after three case runs. The 5\% dispute rate follows closely, while the 20\% run requires 4 cases to break even.
Furthermore, we can see that less than one best case run can more than amortise one worst case run.

% Other Approaches
\begin{table}[tb]
\scriptsize
\centering
\begin{tabular}{|l|l|l|l|llll|}
\hline
\multirow{2}{*}{\textbf{Case}\vspace{-9pt}} &
  \multicolumn{1}{c|}{\multirow{2}{*}{\textbf{Approach}\vspace{-9pt}}} &
  \multirow{2}{*}{\textbf{\begin{tabular}[c]{@{}c@{}}Deploy-\\ ment\vspace{-9pt}\end{tabular}}} &
  \multirow{2}{*}{\textbf{\begin{tabular}[c]{@{}c@{}}Avg. Case\\ Execution\vspace{-9pt}\end{tabular}}} &
  \multicolumn{4}{|c|}{\textbf{Leafhopper Overhead}} \\ \cline{5-8} 
                  &       &  &  & \multicolumn{1}{|l|}{\begin{tabular}[c]{@{}c@{}}Deploy-\\ ment\end{tabular}} & \multicolumn{1}{l|}{{\begin{tabular}[c]{@{}c@{}}Best\\ Case\end{tabular}}} & \multicolumn{1}{l|}{{\begin{tabular}[c]{@{}c@{}}Bad\\ Case\end{tabular}}} & 
                  {\begin{tabular}[c]{@{}c@{}}Worst\\ Case\end{tabular}} \\ \hline
\multirow{3}{*}{\vspace{-7pt}\rotatebox[origin=c]{90}{\begin{tabular}[c]{@{}c@{}}Supply\\Chain\end{tabular}}} & 
{Garc{\'i}a-Ba{\~n}uelos} et al.~\cite{garcia-banuelosOptimizedExecutionBusiness2017a} & 
\multicolumn{1}{r|}{298.564} & 
\multicolumn{1}{r|}{272,186} & 
\multicolumn{1}{|r|}{1.6} &
\multicolumn{1}{r|}{-0.7} & 
\multicolumn{1}{r|}{0.1} & 
\multicolumn{1}{r|}{0.8} \\ \cline{2-8} & 
Caterpillar~\cite{lopez-pintadoCaterpillarBusinessProcess2019a} & 
\multicolumn{1}{r|}{1,100,590} &
\multicolumn{1}{r|}{566,861} &
\multicolumn{1}{|r|}{-0.3} &
\multicolumn{1}{r|}{-0.8} &
\multicolumn{1}{r|}{-0.5} & 
\multicolumn{1}{r|}{-0.1} \\ \cline{2-8} & 
ChorChain~\cite{corradiniEngineeringTrustableAuditable2022a}  & 
\multicolumn{1}{r|}{2,802,543} &
\multicolumn{1}{r|}{1,156,734} &
\multicolumn{1}{|r|}{-0.7} &
\multicolumn{1}{r|}{-0.9} &
\multicolumn{1}{r|}{-0.7} &
\multicolumn{1}{r|}{-0.6} \\ \cline{2-8} & 
CoBuP~\cite{loukilDecentralizedCollaborativeBusiness2021a} &
\multicolumn{1}{r|}{4,832,706} & 
\multicolumn{1}{r|}{254,661} & 
\multicolumn{1}{|r|}{-0.8} & 
\multicolumn{1}{r|}{-0.6} & 
\multicolumn{1}{r|}{0.2} & 
\multicolumn{1}{r|}{0.9} \\ \hline
% Incident Mngmt
\multirow{3}{*}{\vspace{-7pt}\rotatebox[origin=c]{90}{\begin{tabular}[c]{@{}c@{}}Incident\\ Mgmt.\end{tabular}}} & 
{Garc{\'i}a-Ba{\~n}uelos} et al.~\cite{garcia-banuelosOptimizedExecutionBusiness2017a} & 
\multicolumn{1}{r|}{345.743} & 
\multicolumn{1}{r|}{166,345} & 
\multicolumn{1}{|r|}{1.3} &
\multicolumn{1}{r|}{-0.5} & 
\multicolumn{1}{r|}{0.2} & 
\multicolumn{1}{r|}{1.0} \\ \cline{2-8} & 
Caterpillar~\cite{lopez-pintadointerpretedExecutionBusiness2019} & 
\multicolumn{1}{r|}{1,119,803} & 
\multicolumn{1}{r|}{324,420} & 
\multicolumn{1}{|r|}{-0.3} & 
\multicolumn{1}{r|}{-0.7} & 
\multicolumn{1}{r|}{-0.4} &    
\multicolumn{1}{r|}{0} \\ \cline{2-8} & 
ChorChain~\cite{corradiniEngineeringTrustableAuditable2022a} & 
\multicolumn{1}{r|}{3,278,656} & 
\multicolumn{1}{r|}{1,028,505} & 
\multicolumn{1}{|r|}{-0.8} & 
\multicolumn{1}{r|}{-0.9} & 
\multicolumn{1}{r|}{-0.8} & 
\multicolumn{1}{r|}{-0.7} \\ \cline{2-8} & 
CoBuP~\cite{loukilDecentralizedCollaborativeBusiness2021a} &
\multicolumn{1}{r|}{4,639,652} & 
\multicolumn{1}{r|}{249,378} & 
\multicolumn{1}{|r|}{-0.8} & 
\multicolumn{1}{r|}{-0.6} & 
\multicolumn{1}{r|}{-0.2} & 
\multicolumn{1}{r|}{0.3} \\ \hline
\end{tabular}
\caption{Gas cost of Leafhopper in relation to other approaches.}
\label{tab:gas_others}
\vspace{-5pt}
\end{table}

\paragraph{Cost Compared to Related Work.}
In Table~\ref{tab:gas_others}, we compare the cost of Leafhopper to selected approaches from literature by reporting its relative overhead.\footnote{Due to the different feature sets being supported, these approaches incur different gas costs; cost should not be understood as the only yardstick to compare approaches by. Since our approach in this paper is quite different from full on-chain approaches, we find this comparison worthwhile reporting.} 
We chose {Garc{\'i}a-Ba{\~n}uelos} et al.~\cite{garcia-banuelosOptimizedExecutionBusiness2017a} for its efficient implementation technique; Caterpillar~\cite{lopez-pintadoCaterpillarBusinessProcess2019a} for providing the most complete feature set; and the choreography-based approaches ChorChain~\cite{corradiniEngineeringTrustableAuditable2022a} (compiled) and CoBuP~\cite{loukilDecentralizedCollaborativeBusiness2021a} (interpreted).
% We chose {Garc{\'i}a-Ba{\~n}uelos} et al.~\cite{garcia-banuelosOptimizedExecutionBusiness2017a}, as the work providing the most efficient implementation technique; Caterpillar~\cite{lopez-pintadoCaterpillarBusinessProcess2019a}, as work providing the most complete feature set; and ChorChain~\cite{corradiniEngineeringTrustableAuditable2022a} and CoBuP~\cite{loukilDecentralizedCollaborativeBusiness2021a} as choreography-based approaches, where ChorChain is baed on a compiled and CoBuP an interpreted approach. 

The results of this comparison are in line with our above analysis. Leafhopper’s best case provides vast cost improvements (between $1/10$ to $1/2$ of the cost). The medium case overhead ranges from big improvement ($1/10$) to slightly worse (20\% more expensive) depending on the efficiency and feature support of the approach. The worst case cost ranges from double the cost to still considerably ($3/10$) cheaper. Notably, Caterpillar and ChorChain exhibit the highest cost. However, both provide more features. ChorChain additionally implements answer and response patterns and does not implement net reduction and encodes the process state as simple array type, leading to increased cost. This shows the potential of Leafhopper to improve the cost of more complex implementations with its fixed state verification cost.
\subsection{Qualitative Evaluation}
\label{sec:qualitative}
\vspaceSSecAfter
To gain a more holistic understanding of the proposed approach, we now move to a higher level of abstraction and compare the process channel approach to our baseline on-chain approach, which enacts the entire process on-chain, on the basis of relevant quality attributes.
As such, we assess which guarantees the approaches provide, and which assumptions must hold. %, compared to a full on-chain approach. 
Xu et al.~\cite[Chapter~1.4]{xuArchitectureBlockchainApplications2019} identify the main non-functional properties that blockchain provides: immutability, non-repudiation, integrity, transparency, and equal rights. We summarise our assessment in Table 8.2. 
\begin{table}[b]
\scriptsize
  % \vspaceSSecBefore
  \caption{Assumptions and guarantees of on-chain and process channel enactment.}
  \label{tab:guarantees}
  \centering
  \begin{tabular}{lcc}
  \hline
  \textbf{Assumptions} & ~~ On-Chain & Process Channel \\
  \hline
   Blockchain Reliability & \checkmark & \checkmark \\
   At Least One Honest Participant & \xmark & \checkmark \\
   Explicit Role-Binding & \xmark & \checkmark \\
   Participants are Always Available & \xmark & \checkmark \\
   Security of Off-Chain Protocol & \xmark & \checkmark \\
  \hline
  \textbf{Guarantees} & & \\
   \hline
   Integrity & \checkmark & \checkmark$^{\S}$ \\
   Immutability & \checkmark & \checkmark$^{\dagger}$  \\
   Non-Repudiation & \checkmark & \hspace{3.5pt}\checkmark$^{\dagger\star}$ \\
   Transparency & \checkmark & \checkmark$^{\S}$ \\
   Equal Rights & \checkmark & \checkmark$^{\S}$ \\
  \hline
  \multicolumn{3}{l}{} $^{\star}$Stalling is a non-attributable fault ~~~
  $^{\dagger}$Requires storage of proof  \\
  $^{\S}$Requires access to channel & 
\end{tabular}
\end{table}

\paragraph{Assumptions.}
Both approaches assume the reliability of the blockchain. In addition, process channels require at least one honest participant in a channel; otherwise, colluding participants can install arbitrary state. In contrast, an on-chain approach can still rely on other validators in the blockchain network to verify a transaction. Process channels also require explicit role-binding: roles must be bound to (trigger) hostnames and blockchain accounts. This information must be propagated through the channel. Additionally, participants joining the network must deploy trigger components. This inhibits process flexibility. Also, participants in the channel must be always available, they cannot go offline in-between tasks. They must sign transitions and watch for disputes. %However, the automation of a process already requires the operation of PAIS components, even for the on-chain baseline. However, it is an additional burden, and 
Some usage scenarios, e.g., %the integration of end users into the process, 
energy-constrained wireless devices,
may be ruled out by this requirement. Finally, channels introduce additional components and, thus, additional attack surface. % for security threats. %This is additional cost and risk to be considered.

\paragraph{Guarantees.}
As only channel participants see and verify transitions, integrity and transparency can only be demonstrated to participants with access to the channel. Also, equal rights can only hold within the channel. Immutability and non-repudiation require that transition proposals are stored durably by a participant. Otherwise, a faulty participant can submit stale state. Additionally, under certain circumstances a participant can stall the process without being identified as doing such. It is undecidable whether an initiator is stalling the process by not sending the next transition proposal or whether a signing participant has refused to sign it~\cite{mccorryYouSankMy2019}. While the process will be forced to continue on-chain, the channel contract cannot attribute who was at fault, limiting the use of penalties.
%This limits the use of incentives. %of incentive mechanisms.
%A party can not be penalised for being unavailable during a state transition proposal.
%In an on-chain approach, it is always clear which participant is currently responsible for enacting the next task on the blockchain.
%This issue may incentivise unavailability griefing, where a party purposefully triggers a dispute to stall the process temporarily (Coleman alternative?). 
%Non-attributable faults limit the capabilities of incentive mechanisms. As a party can not be penalised for being unavailable during a state transition proposal.

\vspaceSSecBefore
\subsection{Discussion}
\vspaceSSecAfter

In \cref{sec:qualitative}, we analysed the assumptions and guarantees of off-chain enactment. When these assumptions are met, the channel can offer comparable guarantees to an on-chain approach. However, additional complexity is introduced by requiring dedicated channel components, which require a certain degree of redundancy, as they must stay online. While choreographies already require participants to handle messages reliably, process channels increase this reliability requirement. All participants must stay online to advance the choreography and prevent execution forks.
Additionally, process data is no longer durably stored on the blockchain; where such durability is required, it needs to be ensured by the participants off-chain.
Finally, the closed nature of the channel hinders flexibility. Third parties can only verify the honest execution of the process if given access to all messages in the channel. 
% For a process of global interest, a channel is not suitable, especially since a defining cost factor of channels is the number of participants. 

In \cref{sec:quantitative}, we reported on the gas cost of our prototype Leafhopper. While gas costs are an especially limiting factor when enacting processes on public blockchains, they can also limit scalability of private and permissioned deployments, as gas costs are directly linked to the number, computational complexity, and storage requirement of transactions. 
We found that, for processes which are repeated more than 3-4 times without disputes, Leafhopper can significantly reduce the gas cost of on-chain enactment.
In Leafhopper, costs are highly dependent on how often a dispute occurs. If we assume the off-chain protocol will resolve trivial faults, such as temporary connectivity issues, the blockchain will be mainly involved in resolving permanent faults, such as malicious acts or long-term crashes, which constitute a contract breach. For our benchmarked cases, a worst case run was amortised after only one dispute-free run. 
In the case of interorganisational processes, organisations generally form a collaboration to work towards a common business goal; and under typical industry dispute rates, Leafhopper was able to significantly reduce cost. 
Nonetheless, Leafhopper’s cost is case specific and not as predictable as on-chain enactment. This introduces additional uncertainties. We have provided an analysis to help gauge the cost of Leafhopper. Still, future studies are required to address these uncertainties and aspects not covered in our current evaluation, like latency. 
%We expect that Leafhopper will be especially helpful in cases where a high percentage of conformance is expected and non-conforming behaviour would lead to high process costs, e.g., in compliance checking.

\noindent Beyond cost, process channels have the potential to improve 
%the performance in terms of 
further dimensions such as latency and confidentiality. 
On-chain approaches are highly dependent on the underlying latency of the blockchain. Process channels can reduce this reliance. 
%---the time until a transaction can be considered final. 
%Given $n$ participants, Leafhopper requires, for each state transition, 3 off-chain messages. Proposal and confirmation messages can be sent concurrently. In addition, one blockchain transaction is required to submit the final state. Should a dispute be triggered, participants must wait for the dispute window to elapse before the process can be continued on-chain. 
Blockchain latency differs greatly between different blockchain platforms and the required trust assumptions of the application; related work reports latency ranging from seconds (e.g.,~\cite{weberUntrustedBusinessProcess2016a, corradiniModeldrivenEngineeringMultiparty2021a}) to minutes and more (e.g.,~\cite{prybilaRuntimeVerificationBusiness2020a}). Furthermore, blockchain consensus can be highly probabilistic, resulting in high latency outliers (see e.g.,~\cite{prybilaRuntimeVerificationBusiness2020a, corradiniModeldrivenEngineeringMultiparty2021a}). 
%Off-chain messages, on the other hand, only incur the standard network latency. 
Similarly, channels reduce the exposure of data. While confidentiality is breached during a dispute phase, there is potential to design the dispute phase in a confidentiality-preserving manner; for example, by utilising zero-knowledge proofs. While these are usually costly operations, they would only be required in the case of a dispute, making their use viable~\cite{zhang2019zchannel}. 
%Furthermore, this would remove the need for a dispute window. 
There are further topics not addressed and left to future work.

\begin{itemize}
    \item \textit{Length of Dispute Window}: The choice of the dispute window is an important factor. It must be chosen so that an honest participant has time to react to stale state. Thus, it must be multiples of the underlying blockchain latency. There is currently no consensus in literature on how to determine this. In the BPM context, the particular business case may also influence this choice.
    \item \textit{Dispute Phase Design}: In a choreography, a participant that stops to collaborate will stall the process indefinitely. A more advanced dispute design could penalise faulty participants and replace them. There is potential to design dispute processes, specific to business cases, to incentivise honest participation. However, such a design is limited by non-attributable faults.
    \item \textit{Channel Networks:} In channel networks, multiple channels are supported by one \textit{root contract}. In our current design, the channel smart contract is application specific. Exploring a design where a contract can support multiple processes could pave the way toward a network of  cost efficient, blockchain-based choreographies.
\end{itemize}
%
% Alternatively as paragraphs:
%
%\paragraph{Length of Dispute Window}: The choice of the dispute window is an important factor. It must be chosen so that an honest participant has time to react to stale state. Thus, it must be multiples of the underlying blockchain latency. There is currently no consensus in literature on how to determine this. In the BPM context, the particular business case may also influence this choice.

%\paragraph{Dispute Phase Design}: In a choreography, a participant that stops to collaborate will stall the process indefinitely. A more advanced dispute design could penalise faulty participants and replace them. There is potential to design dispute processes, specific to business cases, to incentivise honest participation. However, such a design is limited by non-attributable faults.

%\paragraph{Channel Networks:} In channel networks, multiple channels are supported by one smart contract. In our current design, the contract is application-specific. Exploring a design where a contract can support multiple process could pave the way towards a network of cost efficient, blockchain-based choreographies.
        
    %\item \textbf{Improved State Verification:} In Leafhopper, each participant's signature is verified independently. However, signature compression and public-key aggregation allow multiple signatures to be aggregated, reducing the complexity of verification from $O(n)$, where $n$ is the number of signatures, to $O(1)$~\cite{bonehCompactMultisignaturesSmaller2018}.

%
% \subsubsection{Threats to Validity.} 
%
\vspace{-10pt}
\vspaceSecBefore
\section{Conclusion}
\label{sec:concl}
\vspaceSecAfter

In this paper, we propose to address challenges in inter-organizational process enactment by moving to a layer two approach: blockchain-based state channels.
With this approach, we aim to reduce the on-chain footprint. The quantitative evaluation shows a significant reduction in gas cost for common settings. The qualitative evaluation shows that the blockchain properties largely remain intact when moving to our channel approach---as long as the assumptions are met, such as having at least one honest participant per channel.

Moving communication and state into channels may, in the future, prove useful to achieve lower latency and improved confidentiality---but those aspects were out of scope for this paper, where we focused on the principled approach and and extensive evaluation. 
Future work will, thus, address latency and confidentiality.
Beyond that, we outlined a multitude of other research opportunities, such as the design of process channel networks, where multiple channels are supported by a singular on-chain contract.

%
%In offering these benefits, process channels introduce new assumptions and reduce some of the guarantees offered by blockchain. We have discussed these in Section~\ref{sec:qualitative}. Integrity, transparency, and equal rights only hold within the channel. As a result, third parties interested in the honest execution of the process must join the channel. Additionally, process data is no longer durably stored on the blockchain, but must be stored by channel participants. Finally, non-attributable faults can complicate the dispute resolution.

%
% ---- Bibliography ----
%
% \bibliographystyle{splncs04}
\vspaceSecBefore
\bibliography{bib}
%\printbibliography
%
\end{document}